\documentclass[aps,prl,twocolumn]{revtex4}

\usepackage[normalem]{ulem}
\usepackage{amsmath,amssymb}
\usepackage{graphicx,amsmath,amsfonts}
\usepackage[usenames]{color}
\usepackage{epstopdf}

\newcommand{\be}{\begin{equation}}
\newcommand{\beq}{\begin{eqnarray}}
\newcommand{\eeq}{\end{eqnarray}}

\def \be{\begin{equation}}
\def \ee{\end{equation}}
\def \ba{\begin{array}}
\def \ea{\end{array}}
\def \bea{\begin{eqnarray}}
\def \eea{\end{eqnarray}}

\def \a{{\alpha}}

\def \D{{\Delta}}
\def \d{{\delta}}

\def \ket#1{{\,|\,#1\,\rangle\,}}
\def \bra#1{{\,\langle\,#1\,|\,}}

\begin{document}

\title{Quantized pumping and phase diagram topology of interacting bosons}
\author{Erez Berg$^1$, Michael Levin$^{1,2}$, and Ehud Altman$^3$}
\affiliation{$^1$Department of Physics, Harvard University, Cambridge, MA 02138, USA}
\affiliation{$^2$Department of Physics, University of Maryland, College Park, Maryland
20742-4111, USA}
\affiliation{$^3$Department of Condensed Matter Physics, The Weizmann Institute of
Science, Rehovot, 76100, Israel}
\date{\today }

\begin{abstract}
Interacting lattice bosons at integer filling can support two distinct
insulating phases, which are separated by a critical point: the Mott
insulator and the Haldane insulator\cite{HIprl}. The critical point can be
gapped out by breaking lattice inversion symmetry. Here, we show that
encircling this critical point adiabatically pumps one boson across the
system. When multiple chains are coupled, the two insulating phases are no
longer sharply distinct, but the pumping property survives. This leads to
strict constraints on the topology of the phase diagram of systems of
quasi-one dimensional interacting bosons.
\end{abstract}

\maketitle


In the early 1980s, Thouless \cite{Thouless1983} made the surprising
observation that certain band insulators can sustain dissipationless and
quantized charge transport by adiabatic pumping. The classic example of this
effect is seen in a half filled tight binding chain with two sites per unit
cell\cite{Thouless1983}. As parameters of the hamiltonian are changed
adiabatically along a closed loop around the single gapless point in the two
parameter space, a unit charge is transported through the chain.
This simple observation had interesting implications to other systems. For
example, it was quickly realized \cite{NiuThouless,Avron1985}, that
Laughlin's original argument for the quantization of the Hall conductance
may be formulated in the same mathematical terms as the pumping problem. In
connection with more recent developments, the ideas of topological pumping
through band insulators were precursors of the theoretical\cite%
{Kane2005,Bernevig2006,MooreBalents2007,Fu2007} and subsequent experimental%
\cite{Konig2007,Hasan2008} discovery of topological band insulators. Indeed,
the $Z_2$ topological invariant associated with these systems can be
reformulated in terms of adiabatic pumping 
\cite{Fu2006}.

Although quantized pumping has been discussed primarily in the context of
non--interacting fermions, the concept is much more general.
The pumped charge can be formulated in terms of a topological Chern number
associated with parallel transport of the many-body wave-function in Hilbert
space\cite{NiuThouless,Avron1985}. In particular this formulation ensures
robustness of the quantization to disorder and interaction and also enables
direct extension of the concepts to spin pumping in spin-$1/2$ chains\cite%
{Shindou2005}. All these extensions are adiabatically connected to the case
of a band insulator, either directly or via a Jordan-Wigner transformation.


\begin{figure}[t]
\begin{center}
\includegraphics[width=2.5in]{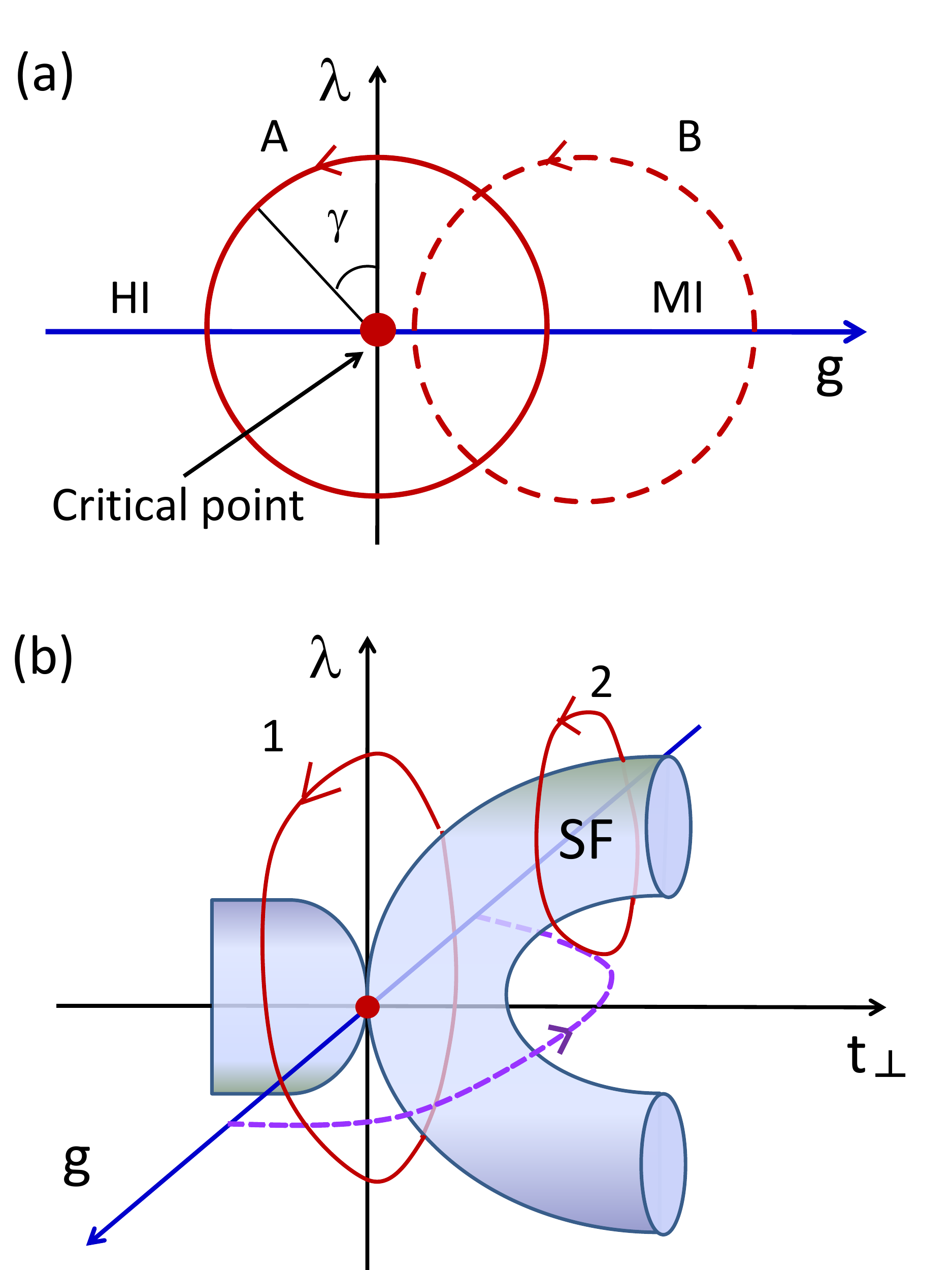}
\end{center}
\par
\vspace{-0.3in} \caption{\emph{Phase diagram topology.} (a) Phase
diagram of a single chain. The parameter $g$ tunes across the
Haldane (HI) to Mott (MI) insulator transition. These two phases
are sharply defined only in presence of inversion symmetry
($\protect\lambda=0$). A closed adiabatic path which encircles the
critical point (path $A$) entails pumping of a single boson across
the insulator. (b) Schematic phase diagram of two coupled chains.
The MI and HI phases can be adiabatically connected via the dashed
path. However, since path 1 pumps one boson per chain, it cannot
be collapsed adiabatically to a point without crossing the gapless
region. Path 2 entails pumping of one boson per two chains.}
\label{fig:pd}
\end{figure}

In this paper we show that a natural model of interacting lattice bosons at
integer filling, which is not directly mappable to a band insulator, allows
quantized transport through Mott insulating states by adiabatic pumping. The
existence of non trivial loops in the gapped regions of parameter space
defines a topological index, which may be associated with the gapless
(superfluid) phases they surround. It also sets constraints on the structure
of the phase diagram, or more precisely, on the topology of the gapless
regions within it.

The basic model we consider is an extended Bose Hubbard model (EBHM), at
integer filling, on coupled chains
\begin{equation}
H=\sum_\a \left[ H_\a+H_{\lambda,{\alpha}}+ H_{\perp,{\alpha}} \right] \text{%
,}  \label{Hmic}
\end{equation}
where
\begin{eqnarray}
H_\a&=& \sum_{j}\left[-t (b_{{\alpha},j}^\dagger b^{\vphantom{\dagger}}_{{%
\alpha},j+1} +\mathrm{H.c.})+\frac{U}{2}n_{{\alpha},j}(n_{{\alpha},j}+1) %
\right]  \notag \\
&+&V \sum_j n_{{\alpha},j}n_{{\alpha},j+1}\text{,}  \label{Hchain}
\end{eqnarray}
is a single chain hamiltonian defined on chain ${\alpha}$. $b^\dagger_{{%
\alpha},j}$ creates a boson at position $j$ in chain ${\alpha}$, and $n_{{%
\alpha},j}\equiv b^\dagger_{{\alpha},j} b^{\vphantom{\dagger}}_{{\alpha},j}$%
. The hamiltonian
\begin{equation}
H_{\lambda,{\alpha}} = \lambda\sum_j\left[ n_{{\alpha},j} b^\dagger_{{\alpha}%
,j} b^{\vphantom{\dagger}}_{{\alpha},j+1} - n_{{\alpha},j+1} b^\dagger_{{%
\alpha},j+1} b^{\vphantom{\dagger}}_{{\alpha},j} + \mathrm{H.c.} \right]
\text{,}  \label{Hlambda}
\end{equation}
is a perturbation that breaks the bond-centered inversion symmetry
of $H_\a$. Finally,
\begin{equation}
H_{\perp,{\alpha}} = \sum_j [-t_\perp (b_{{\alpha},j}^\dagger b^{%
\vphantom{\dagger}}_{{\alpha}+1,j}+\mathrm{H.c.}) + V_\perp n_{{\alpha},j}
n_{{\alpha}+1,j}]  \label{Hperp}
\end{equation}
denotes interchain coupling. The model (\ref{Hmic}) or related hamiltonians
can be realized with ultracold dipolar molecules or atoms with optically
induced dipole moments\cite{Pupillo2010}. Crucial for our analysis is the
presence of the perturbation $\lambda$, which breaks the inversion symmetry
of the chain. It will be naturally generated if the underlying optical
potential is not symmetric under inversion. Such a lattice potential can be
produced by two lasers, one of which has double the wavelength of the other.
In one extreme limit this configuration gives rise to a lattice of double
well potentials\cite{PortoDouble,Folling2007}, which indeed are not
inversion symmetric in general.

\emph{A single chain --} We have shown previously, that the EBHM on a single
chain (Eq. \ref{Hchain}) exhibits a quantum phase transition from a Mott
insulating (MI) state to a novel gapped phase, which we termed a ``Haldane
insulator" (HI), upon increasing the nearest neighbor interaction\cite{HIprl}%
. Both phases are completely disordered in the sense that they do not break
any symmetry of the Hamiltonian. The new state is 
analogous to the Haldane gapped state of spin-1 chains, and 
is characterized by a string order parameter, albeit in the boson density
rather than the spin. It was later shown \cite%
{HIprb,Gu2009,Pollmann2009,Pollmann2010}, that the distinction between the
HI and MI phases is protected by lattice inversion symmetry. A perturbation,
such as $H_{\lambda,{\alpha}}$ above, which breaks the inversion symmetry
about a bond, opens a gap at the HI--MI transition and allows adiabatic
connection between the two gapped phases. Thus, in the two parameter space $%
(V,\lambda)$ the transition becomes an isolated critical point. We shall
argue that an adiabatic passage around the critical point entails transport
of a single boson through the chain.

To see this we turn to the long wavelength description of the extended
Hubbard chain, with the inversion symmetry breaking perturbation $\lambda$.
Near to the HI--MI phase transition it is given by the following Sine-Gordon
field theory\cite{HIprb}
\begin{eqnarray}
H_0 &=&\frac{u}{2\pi }\int dx\Big[ K\left( \partial _{x}\theta \right) ^{2}+%
\frac{1}{K}\left( \partial_{x}\phi\right) ^{2}  \notag \\
&&-g\cos(2\phi)-\lambda\sin(2\phi)\Big]\text{,}  \label{Hp}
\end{eqnarray}
with the Luttinger parameter $K$ in the regime $1/2<K<2$. The parameter $g$
is in general a complicated function of the microscopic interactions. $%
g(U,V,t)>0$ in the MI, $g<0$ in HI and vanishes on the critical line
separating these two phases. A naive continuum limit gives the approximate
dependence $g\approx U/2-V$~\cite{HIprb}. Here $\partial_x\phi/\pi=\rho$ is
the long-wavelength component of the boson density, $\theta$ is its dual
field satisfying $[\partial_x\phi(x),\theta(x^{\prime})]=i\pi\delta(x-x^{%
\prime})$, and $u$ is the sound velocity. Note that under inversion, $%
\rho(x)\rightarrow \rho(-x)$, therefore $\phi(x)\rightarrow -\phi(-x)$,
which makes it clear that the $\lambda$ term is odd under inversion. The
last two terms can be written compactly as $\tilde{g}\cos(2\phi-\chi)$,
where $\tilde{g}=\sqrt{g^2+\lambda^2}$ and $\chi=\arctan(\lambda/g)$. In the
regime of interest 
$K<2$, making the cosine term relevant. $\cos(2n\phi)$ and
$\sin(2n\phi)$ with $n>1$ may also appear in $H_0$, but we assume
that these terms are irrelevant at the HI--MI
critical point, $(\lambda=0,g=0)$. 

The critical point is entirely surrounded by a gapped state (see Fig. \ref%
{fig:pd}a) in which the field $\phi$ is essentially locked to the value $%
\chi/2$. Therefore an adiabatic change of the system parameters, which takes
it in a counter clockwise loop around the critical point incurs a continuous
change of $\phi(x)$ by $\pi$ everywhere in space. By definition of the
field, $\phi(x)$ suffers a $\pi$ shift every time a particle passes through $%
x$. The last observation implies the transport of exactly one
boson from left to right in a counter-clockwise loop. Another way
to derive the quantization is to refermionize the field theory
(\ref{Hp}). The quantized charge can be computed directly for
$K=1$, which maps to free fermions \cite{supplementary}. It
follows for other values of $K$ by adiabatic continuity.

To enable continuous pumping, the chain must be connected to
gapless reservoirs. This arises naturally in a realization using
an optical lattice and a harmonic trap in which the incompressible
phase will be flanked by superfluid wings. The adiabaticity
condition needed to ensure quantized pumping is $\dot\chi\ll\D\sim
\Lambda ({\tilde g}/\Lambda)^{1/(2-K)}$\cite{supplementary}, where
$\D$ is the gap along the cycle. $\Lambda$, the ultraviolet cutoff
of the continuum theory is of the order of the bandwidth $2t$.



The topological character of the pumped charge makes it robust to small
perturbations of the hamiltonian\cite{NiuThouless}. In particular, for the
case of many weakly coupled chains, driving all chains adiabatically along
loop $A$ still pumps one boson per chain. For arbitrary coupling between
chains, we shall see that the quantization of the pumped charge imposes
stringent constraints on the topology of the phase diagram in the enlarged
parameter space. We demonstrate this below using the example of two coupled
chains and then comment on generalizations to any number of coupled chains.


\emph{Two coupled chains.} The critical point at $(g,\lambda)=0$ is unstable
to weak tunnel coupling $t_\perp$ between a pair of chains\cite{HIprb}.
Using an RG analysis we have shown that the critical point expands to a
gapless phase (Luttinger liquid) with radius $\sim t_\perp^{\eta}$ around
the origin in the space $(g,\lambda)$, where the precise exponent $\eta$ is
given in ref. [\onlinecite{HIprb}].

How is the pumped charge associated with an adiabatic cycle around the
critical point, affected by turning on the inter-chain coupling $t_\perp$?
As long as the path encircles the gapless region from the outside, then it
is adiabatically connected to the non--trivial pumping cycle around the
HI--MI critical point of the decoupled chains. The topological Chern number
cannot change and hence the pumped charge must remain quantized at one boson
per chain upon encircling the gapless region. Below we address the evolution
of the gapless region for increasing inter--chain coupling beyond
weak-coupling.

\begin{figure}[t]
\begin{center}
\includegraphics[width=2.4in]{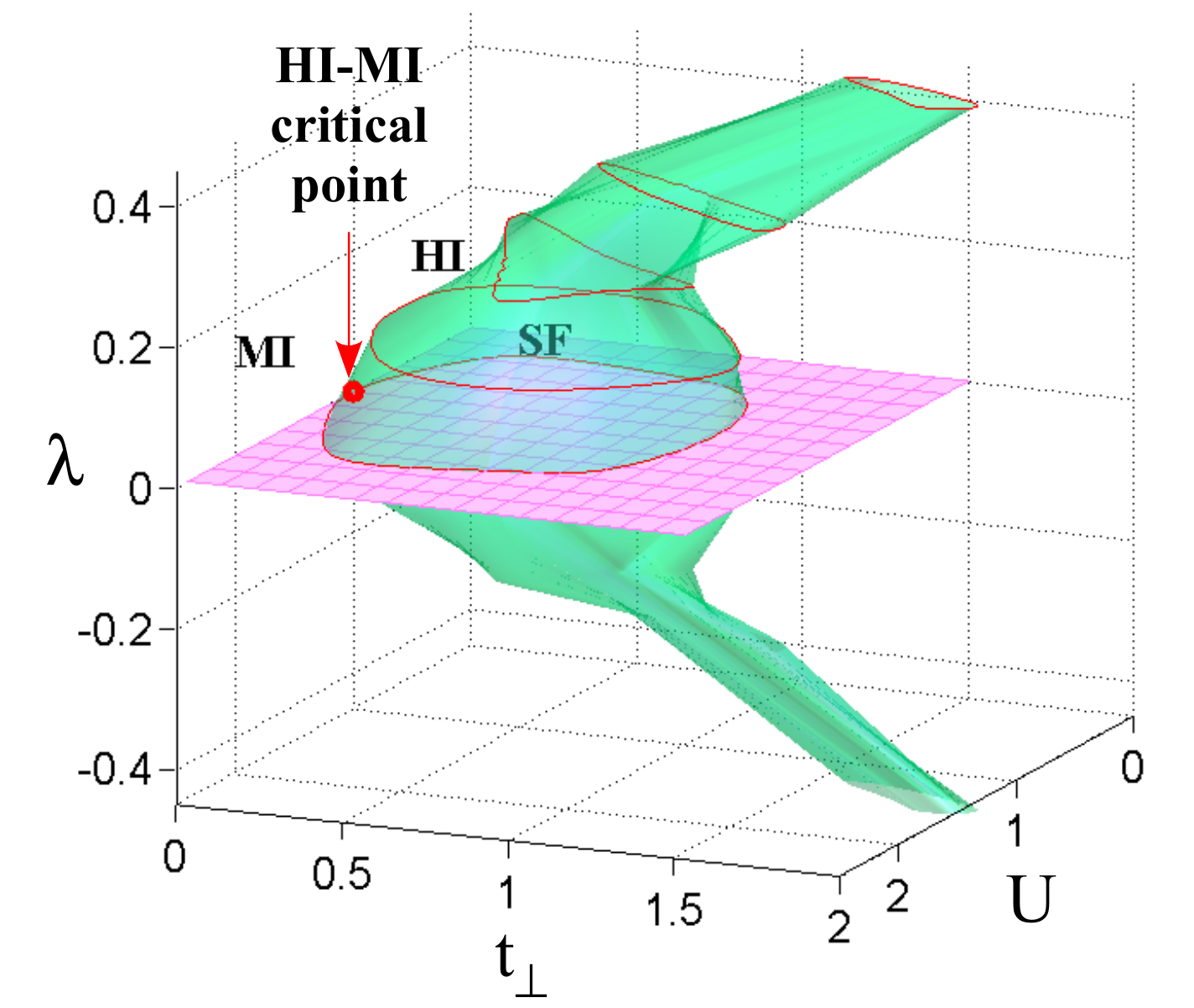}
\end{center}
\par
\vspace{-.3in}
\caption{\emph{Phase diagram of the spin-1 two--leg ladder} defined in Eq. (%
\protect\ref{Hspin}) as a function of $U$, $t_\perp$, and
$\protect\lambda$, calculated using DMRG. We have fixed
$V_\perp=2t_\perp$. For $t_\perp=\protect\lambda=0$, there are two
distinct gapped phases, the HI and MI, which are separated by a
critical point at $U\approx 1$. Upon turning on $t_\perp$, the
critical point expands to a finite superfluid (SF) region, and the
two gapped phases are not sharply distinct. For
$\protect\lambda>0$, the MI--HI critical point at $t_\perp=0$
becomes gapped. The gapless region shrinks upon increasing
$\protect\lambda$, but does not disappear. The phase diagram has
the same topology as in Fig. \protect\ref{fig:pd}b.}
\label{fig:dmrg}
\end{figure}

To understand how the phase diagram evolves with stronger values of
inter-chain coupling we should take into account another crucial fact. For
two chains there is no sharp distinction between the HI and MI phases, even
in the presence of inversion symmetry ($\lambda=0$)\cite{Pollmann2009}. This
means that the HI and MI phases of two decoupled chains can be connected
adiabatically by a path in hamiltonian space going through a region with non
zero inter--chain coupling. We demonstrate this explicitly using a Density
Matrix Renormalization Group (DMRG) calculation of the following spin-1
ladder model:
\begin{eqnarray}
H_{spin}&=& \sum_{i,{\alpha}}\Big[V S^z_{{\alpha}, i} S^z_{{\alpha},i+1}-
t(S^+_{{\alpha}, i} S^-_{{\alpha},i+1}+\mathrm{H.c.})+U(S^z_{{\alpha},i})^2
\notag \\
&+&\lambda (S^z_{{\alpha},i} S^+_{{\alpha},i} S^-_{{\alpha},i+1}-S^z_{{\alpha%
},i+1} S^+_{{\alpha},i+1} S^-_{{\alpha},i}+\mathrm{H.c.}) \Big]  \notag \\
&+&\sum_i \Big[V_\perp S^z_{1, i} S^z_{2,i}-t_\perp(S^+_{1, i} S^-_{2,i}+%
\mathrm{H.c.})\Big]  \label{Hspin}
\end{eqnarray}
This model can be thought of as a truncation of the EBHM (\ref{Hmic}) to the
space of the three lowest occupation states $n_i=S^z_i+1$\cite{Altman2002}.
Crucially, the two models have the same low energy limit\cite%
{Schulz1986,HIprb}.

Fig. 2 shows the phase diagram of the model (\ref{Hspin}), as a function of $%
t_\perp$, $\lambda$ and $U$, which is used to tune the MI--HI transition. ($%
U $ is related to $g$ in Eq. \ref{Hp} by $g\propto U-U_c$, where $U_c$ is
the location of the MI--HI transition.) We have fixed $V=2t$ and $%
V_\perp=2t_\perp$. The phase diagram was determined by measuring the spin
gap, $\Delta_s=E(S=1)-E(S=0)$, and extrapolating it to the thermodynamic
limit. System sizes of up to $L=64\times 2$ were used, keeping $m=200$
states.

We see in Fig. \ref{fig:dmrg}, that upon increasing the
inter-chain coupling $t_\perp$ and $V_\perp$ the HI-MI critical
point first expands to a gapless region as predicted by the weak
coupling theory \cite{HIprb}, but collapses at stronger coupling,
allowing for an adiabatic connection between the HI and MI states.
The fact that the gapless region ends may at first seem
contradictory to our previous assertion that an adiabatic loop
around this region in the space $(g,\lambda)$ entails pumping of
one boson per chain. If the gapless region ends, and the loop can
be collapsed adiabatically to a trivial point in the gapped state
with increasing inter-chain coupling, how can it sustain a
non-trivial Chern number?

To avoid this contradiction, the gapless region must split into two branches
in the $\pm\lambda$ directions, 
which either extend indefinitely, or terminate discontinuously on a 1$^{st}$
order transition plane. With this topology, a loop surrounding the original
critical point at $t_\perp=0$ cannot be collapsed adiabatically into a
point. The numerically obtained phase diagram in Fig. \ref{fig:dmrg} is
consistent with these considerations: although the superfluid region in the $%
\lambda=0$ plane is finite, it has two branches which extend in
the $\pm \lambda$ directions. These branches do not terminate up
to the largest values of $\lambda$ we examined (in \cite{supplementary} we present results for higher $\lambda$ values).

Given the topology of the gapless phase it is natural to ask what is the
pumped charge associated with a path surrounding only one of the two
branches at either positive or negative $\lambda$ (path 2 in Fig. \ref%
{fig:pd}b). Such a path has no counterpart in the single chain system and it
cannot be continuously deformed into a loop that surrounds an isolated
critical point. Nevertheless, we argue that the Chern number associated with
this path is determined by the topological character of the HI-MI critical
point. A simple way to approach this problem is to note that two loops, each
encircling one of the two branches, can be deformed into a single loop which
encircles both branches. Such a loop corresponds to pumping of two bosons as
discussed above. Therefore, by symmetry of the $\pm\lambda$ branches, each
of the isolated loops entails pumping of one boson along the ladder, or half
a boson per chain.

The distribution of quantized charge among different loops in parameter
space can be succinctly represented in terms of a fictitious quantized
magnetic flux running through gapless regions in the three dimensional
parameter space. Two flux quanta, one for each chain, are inserted through
the isolated HI-MI critical point in the $t_\perp$ direction, and must split
evenly between the two branches at $\pm\lambda$. The quantized pumping
therefore defines a topological index, the fictitious quantized-flux, that
is associated with the gapless phases.

\emph{More than two chains --} Without inter-chain coupling, $N$ parallel
chains are just $N$ copies of the single chain problem, and so an adiabatic
cycle around the critical point implies pumping of $N$ bosons along the
decoupled chains. As before, this charge cannot change suddenly with the
introduction of inter-chain coupling $t_\perp$. Hence, in the extended
parameter space the critical point at the origin $t_\perp=0$ is a source of $%
N$ quanta of the fictitious flux. The gapless phase at finite $t_\perp$ may
branch out, as in the case of two chains, while the fictitious flux running
through all the branches must add up to exactly $N$.

There is another topological constraint on the branching of the gapless
phase with increasing $t_\perp$. From the construction of Ref. \cite%
{Pollmann2009,Pollmann2010}, follows a sharp
distinction between the Haldane insulator and the Mott insulator
phase on any ladder with an odd number of chains. That is, without
breaking inversion symmetry, $2N+1$ decoupled chains in the HI
phase cannot be connected adiabatically to
decoupled chains in the MI by an adiabatic path going through finite $%
t_\perp $, in contrast to the two leg case considered above.
Therefore in a ladder with odd number of legs the gapless phase must persist
indefinitely on the plane with inversion symmetry, i.e $\lambda=0$.


\emph{Conclusions--} Topological properties of matter are usually associated
with gapped regions of the phase diagram. Here, we have shown that in a
model of interacting bosons, it is natural to associate a topological
``flux'' with the gapless (superfluid) regions, which is defined by the
pumped charge upon encircling these regions adiabatically. This property can
be argued to be more profound, in the sense that the gapped phases discussed
here are only distinct from each other as long as certain symmetries (\emph{%
e.g.} inversion symmetry) are preserved, while the topological flux
associated with the gapless region is robust against arbitrary particle
number conserving perturbations. This principle can be used to impose
constraints on the topology of the phase diagram; for example, it implies
that a gapless region which carries a non-zero topological flux cannot
terminate.

Similarly, topological insulators in two and three dimensions are only
well-defined as long as time-reversal symmetry is preserved. However, the
gapless region separating the topologically trivial and non-trivial phase
may carry a topological ``flux'', which remains well-defined even when
time-reversal symmetry is broken. That can hopefully shed new light on the
nature of topological insulators.\cite{Zohar}

\emph{Acknowledgements.} This work was supported in part by NSF
under grants DMR-0705472 and DMR-0757145 (EB), The US Israel BSF
(EA and EB), ISF (EA), and the Harvard Society of Fellows (ML).
The calculations were run on the Odyssey cluster supported by the
FAS Sciences Division Research Computing Group. EA and EB thank
the Aspen Center for Physics, where part of this work was done.

\bibliography{pumping}

\begin{thebibliography}{25}
\expandafter\ifx\csname natexlab\endcsname\relax\def\natexlab#1{#1}\fi
\expandafter\ifx\csname bibnamefont\endcsname\relax
  \def\bibnamefont#1{#1}\fi
\expandafter\ifx\csname bibfnamefont\endcsname\relax
  \def\bibfnamefont#1{#1}\fi
\expandafter\ifx\csname citenamefont\endcsname\relax
  \def\citenamefont#1{#1}\fi
\expandafter\ifx\csname url\endcsname\relax
  \def\url#1{\texttt{#1}}\fi
\expandafter\ifx\csname urlprefix\endcsname\relax\def\urlprefix{URL }\fi
\providecommand{\bibinfo}[2]{#2}
\providecommand{\eprint}[2][]{\url{#2}}

\bibitem[{\citenamefont{Dalla~Torre et~al.}(2006)\citenamefont{Dalla~Torre,
  Berg, and Altman}}]{HIprl}
\bibinfo{author}{\bibfnamefont{E.~G.} \bibnamefont{Dalla~Torre}},
  \bibinfo{author}{\bibfnamefont{E.}~\bibnamefont{Berg}}, \bibnamefont{and}
  \bibinfo{author}{\bibfnamefont{E.}~\bibnamefont{Altman}},
  \bibinfo{journal}{Phys. Rev. Lett.} \textbf{\bibinfo{volume}{97}},
  \bibinfo{pages}{260401} (\bibinfo{year}{2006}).

\bibitem[{\citenamefont{Thouless}(1983)}]{Thouless1983}
\bibinfo{author}{\bibfnamefont{D.~J.} \bibnamefont{Thouless}},
  \bibinfo{journal}{Phys. Rev. B} \textbf{\bibinfo{volume}{27}},
  \bibinfo{pages}{6083} (\bibinfo{year}{1983}).

\bibitem[{\citenamefont{Niu and Thouless}(1984)}]{NiuThouless}
\bibinfo{author}{\bibfnamefont{Q.}~\bibnamefont{Niu}} \bibnamefont{and}
  \bibinfo{author}{\bibfnamefont{D.~J.} \bibnamefont{Thouless}},
  \bibinfo{journal}{Journal of Physics A: Mathematical and General}
  \textbf{\bibinfo{volume}{17}}, \bibinfo{pages}{2453} (\bibinfo{year}{1984}).

\bibitem[{\citenamefont{Avron and Seiler}(1985)}]{Avron1985}
\bibinfo{author}{\bibfnamefont{J.~E.} \bibnamefont{Avron}} \bibnamefont{and}
  \bibinfo{author}{\bibfnamefont{R.}~\bibnamefont{Seiler}},
  \bibinfo{journal}{Phys. Rev. Lett.} \textbf{\bibinfo{volume}{54}},
  \bibinfo{pages}{259} (\bibinfo{year}{1985}).

\bibitem[{\citenamefont{Kane and Mele}(2005)}]{Kane2005}
\bibinfo{author}{\bibfnamefont{C.~L.} \bibnamefont{Kane}} \bibnamefont{and}
  \bibinfo{author}{\bibfnamefont{E.~J.} \bibnamefont{Mele}},
  \bibinfo{journal}{Phys. Rev. Lett.} \textbf{\bibinfo{volume}{95}},
  \bibinfo{pages}{146802} (\bibinfo{year}{2005}).

\bibitem[{\citenamefont{Bernevig et~al.}(2006)\citenamefont{Bernevig, Hughes,
  and Zhang}}]{Bernevig2006}
\bibinfo{author}{\bibfnamefont{B.~A.} \bibnamefont{Bernevig}},
  \bibinfo{author}{\bibfnamefont{T.~L.} \bibnamefont{Hughes}},
  \bibnamefont{and} \bibinfo{author}{\bibfnamefont{S.-C.} \bibnamefont{Zhang}},
  \bibinfo{journal}{Science} \textbf{\bibinfo{volume}{314}},
  \bibinfo{pages}{1757} (\bibinfo{year}{2006}).

\bibitem[{\citenamefont{Moore and Balents}(2007)}]{MooreBalents2007}
\bibinfo{author}{\bibfnamefont{J.~E.} \bibnamefont{Moore}} \bibnamefont{and}
  \bibinfo{author}{\bibfnamefont{L.}~\bibnamefont{Balents}},
  \bibinfo{journal}{Phys. Rev. B} \textbf{\bibinfo{volume}{75}},
  \bibinfo{pages}{121306} (\bibinfo{year}{2007}).

\bibitem[{\citenamefont{Fu et~al.}(2007)\citenamefont{Fu, Kane, and
  Mele}}]{Fu2007}
\bibinfo{author}{\bibfnamefont{L.}~\bibnamefont{Fu}},
  \bibinfo{author}{\bibfnamefont{C.~L.} \bibnamefont{Kane}}, \bibnamefont{and}
  \bibinfo{author}{\bibfnamefont{E.~J.} \bibnamefont{Mele}},
  \bibinfo{journal}{Phys. Rev. Lett.} \textbf{\bibinfo{volume}{98}},
  \bibinfo{pages}{106803} (\bibinfo{year}{2007}).

\bibitem[{\citenamefont{Konig et~al.}(2007)\citenamefont{Konig, Wiedmann,
  Brune, Roth, Buhmann, Molenkamp, Qi, and Zhang}}]{Konig2007}
\bibinfo{author}{\bibfnamefont{M.}~\bibnamefont{Konig}},
  \bibinfo{author}{\bibfnamefont{S.}~\bibnamefont{Wiedmann}},
  \bibinfo{author}{\bibfnamefont{C.}~\bibnamefont{Brune}},
  \bibinfo{author}{\bibfnamefont{A.}~\bibnamefont{Roth}},
  \bibinfo{author}{\bibfnamefont{H.}~\bibnamefont{Buhmann}},
  \bibinfo{author}{\bibfnamefont{L.~W.} \bibnamefont{Molenkamp}},
  \bibinfo{author}{\bibfnamefont{X.-L.} \bibnamefont{Qi}}, \bibnamefont{and}
  \bibinfo{author}{\bibfnamefont{S.-C.} \bibnamefont{Zhang}},
  \bibinfo{journal}{Science} \textbf{\bibinfo{volume}{318}},
  \bibinfo{pages}{766} (\bibinfo{year}{2007}).

\bibitem[{\citenamefont{{Hsieh} et~al.}(2008)\citenamefont{{Hsieh}, {Qian},
  {Wray}, {Xia}, {Hor}, {Cava}, and {Hasan}}}]{Hasan2008}
\bibinfo{author}{\bibfnamefont{D.}~\bibnamefont{{Hsieh}}},
  \bibinfo{author}{\bibfnamefont{D.}~\bibnamefont{{Qian}}},
  \bibinfo{author}{\bibfnamefont{L.}~\bibnamefont{{Wray}}},
  \bibinfo{author}{\bibfnamefont{Y.}~\bibnamefont{{Xia}}},
  \bibinfo{author}{\bibfnamefont{Y.~S.} \bibnamefont{{Hor}}},
  \bibinfo{author}{\bibfnamefont{R.~J.} \bibnamefont{{Cava}}},
  \bibnamefont{and} \bibinfo{author}{\bibfnamefont{M.~Z.}
  \bibnamefont{{Hasan}}}, \bibinfo{journal}{\nat}
  \textbf{\bibinfo{volume}{452}}, \bibinfo{pages}{970} (\bibinfo{year}{2008}),
  \eprint{0902.1356}.

\bibitem[{\citenamefont{Fu and Kane}(2006)}]{Fu2006}
\bibinfo{author}{\bibfnamefont{L.}~\bibnamefont{Fu}} \bibnamefont{and}
  \bibinfo{author}{\bibfnamefont{C.~L.} \bibnamefont{Kane}},
  \bibinfo{journal}{Phys. Rev. B} \textbf{\bibinfo{volume}{74}},
  \bibinfo{pages}{195312} (\bibinfo{year}{2006}).

\bibitem[{\citenamefont{{Shindou}}(2005)}]{Shindou2005}
\bibinfo{author}{\bibfnamefont{R.}~\bibnamefont{{Shindou}}},
  \bibinfo{journal}{Journal of the Physical Society of Japan}
  \textbf{\bibinfo{volume}{74}}, \bibinfo{pages}{1214} (\bibinfo{year}{2005}),
  \eprint{arXiv:cond-mat/0312668}.

\bibitem[{\citenamefont{Pupillo et~al.}(2010)\citenamefont{Pupillo, Micheli,
  Boninsegni, Lesanovsky, and Zoller}}]{Pupillo2010}
\bibinfo{author}{\bibfnamefont{G.}~\bibnamefont{Pupillo}},
  \bibinfo{author}{\bibfnamefont{A.}~\bibnamefont{Micheli}},
  \bibinfo{author}{\bibfnamefont{M.}~\bibnamefont{Boninsegni}},
  \bibinfo{author}{\bibfnamefont{I.}~\bibnamefont{Lesanovsky}},
  \bibnamefont{and} \bibinfo{author}{\bibfnamefont{P.}~\bibnamefont{Zoller}},
  \bibinfo{journal}{Phys. Rev. Lett.} \textbf{\bibinfo{volume}{104}},
  \bibinfo{pages}{223002} (\bibinfo{year}{2010}).

\bibitem[{\citenamefont{Sebby-Strabley
  et~al.}(2006)\citenamefont{Sebby-Strabley, Anderlini, Jessen, and
  Porto}}]{PortoDouble}
\bibinfo{author}{\bibfnamefont{J.}~\bibnamefont{Sebby-Strabley}},
  \bibinfo{author}{\bibfnamefont{M.}~\bibnamefont{Anderlini}},
  \bibinfo{author}{\bibfnamefont{P.~S.} \bibnamefont{Jessen}},
  \bibnamefont{and} \bibinfo{author}{\bibfnamefont{J.~V.} \bibnamefont{Porto}},
  \bibinfo{journal}{Phys. Rev. A} \textbf{\bibinfo{volume}{73}},
  \bibinfo{pages}{033605} (\bibinfo{year}{2006}).

\bibitem[{\citenamefont{{F{\"o}lling} et~al.}(2007)\citenamefont{{F{\"o}lling},
  {Trotzky}, {Cheinet}, {Feld}, {Saers}, {Widera}, {M{\"u}ller}, and
  {Bloch}}}]{Folling2007}
\bibinfo{author}{\bibfnamefont{S.}~\bibnamefont{{F{\"o}lling}}},
  \bibinfo{author}{\bibfnamefont{S.}~\bibnamefont{{Trotzky}}},
  \bibinfo{author}{\bibfnamefont{P.}~\bibnamefont{{Cheinet}}},
  \bibinfo{author}{\bibfnamefont{M.}~\bibnamefont{{Feld}}},
  \bibinfo{author}{\bibfnamefont{R.}~\bibnamefont{{Saers}}},
  \bibinfo{author}{\bibfnamefont{A.}~\bibnamefont{{Widera}}},
  \bibinfo{author}{\bibfnamefont{T.}~\bibnamefont{{M{\"u}ller}}},
  \bibnamefont{and} \bibinfo{author}{\bibfnamefont{I.}~\bibnamefont{{Bloch}}},
  \bibinfo{journal}{\nat} \textbf{\bibinfo{volume}{448}}, \bibinfo{pages}{1029}
  (\bibinfo{year}{2007}), \eprint{0707.3985}.

\bibitem[{\citenamefont{Berg et~al.}(2008)\citenamefont{Berg, Dalla~Torre,
  Giamarchi, and Altman}}]{HIprb}
\bibinfo{author}{\bibfnamefont{E.}~\bibnamefont{Berg}},
  \bibinfo{author}{\bibfnamefont{E.~G.} \bibnamefont{Dalla~Torre}},
  \bibinfo{author}{\bibfnamefont{T.}~\bibnamefont{Giamarchi}},
  \bibnamefont{and} \bibinfo{author}{\bibfnamefont{E.}~\bibnamefont{Altman}},
  \bibinfo{journal}{Phys. Rev. B} \textbf{\bibinfo{volume}{77}},
  \bibinfo{pages}{245119} (\bibinfo{year}{2008}).

\bibitem[{\citenamefont{Gu and Wen}(2009)}]{Gu2009}
\bibinfo{author}{\bibfnamefont{Z.-C.} \bibnamefont{Gu}} \bibnamefont{and}
  \bibinfo{author}{\bibfnamefont{X.-G.} \bibnamefont{Wen}},
  \bibinfo{journal}{Phys. Rev. B} \textbf{\bibinfo{volume}{80}},
  \bibinfo{pages}{155131} (\bibinfo{year}{2009}).

\bibitem[{\citenamefont{{Pollmann} et~al.}(2009)\citenamefont{{Pollmann},
  {Berg}, {Turner}, and {Oshikawa}}}]{Pollmann2009}
\bibinfo{author}{\bibfnamefont{F.}~\bibnamefont{{Pollmann}}},
  \bibinfo{author}{\bibfnamefont{E.}~\bibnamefont{{Berg}}},
  \bibinfo{author}{\bibfnamefont{A.~M.} \bibnamefont{{Turner}}},
  \bibnamefont{and}
  \bibinfo{author}{\bibfnamefont{M.}~\bibnamefont{{Oshikawa}}},
  \bibinfo{journal}{ArXiv e-prints}  (\bibinfo{year}{2009}),
  \eprint{0909.4059}.

\bibitem[{\citenamefont{{Pollmann} et~al.}(2010)\citenamefont{{Pollmann},
  {Turner}, {Berg}, and {Oshikawa}}}]{Pollmann2010}
\bibinfo{author}{\bibfnamefont{F.}~\bibnamefont{{Pollmann}}},
  \bibinfo{author}{\bibfnamefont{A.~M.} \bibnamefont{{Turner}}},
  \bibinfo{author}{\bibfnamefont{E.}~\bibnamefont{{Berg}}}, \bibnamefont{and}
  \bibinfo{author}{\bibfnamefont{M.}~\bibnamefont{{Oshikawa}}},
  \bibinfo{journal}{Phys. Rev. B} \textbf{\bibinfo{volume}{81}},
  \bibinfo{pages}{064439} (\bibinfo{year}{2010}).

\bibitem[{sup()}]{supplementary}
\bibinfo{note}{See supplementary online material.}

\bibitem[{\citenamefont{Altman and Auerbach}(2002)}]{Altman2002}
\bibinfo{author}{\bibfnamefont{E.}~\bibnamefont{Altman}} \bibnamefont{and}
  \bibinfo{author}{\bibfnamefont{A.}~\bibnamefont{Auerbach}},
  \bibinfo{journal}{Phys. Rev. Lett.} \textbf{\bibinfo{volume}{89}},
  \bibinfo{pages}{250404} (\bibinfo{year}{2002}).

\bibitem[{\citenamefont{Schulz}(1986)}]{Schulz1986}
\bibinfo{author}{\bibfnamefont{H.~J.} \bibnamefont{Schulz}},
  \bibinfo{journal}{Phys. Rev. B} \textbf{\bibinfo{volume}{34}},
  \bibinfo{pages}{6372} (\bibinfo{year}{1986}).

\bibitem[{\citenamefont{Ringel and Altman}()}]{Zohar}
\bibinfo{author}{\bibfnamefont{Z.}~\bibnamefont{Ringel}} \bibnamefont{and}
  \bibinfo{author}{\bibfnamefont{E.}~\bibnamefont{Altman}},
  \bibinfo{note}{unpublished}.

\bibitem[{\citenamefont{{Luther} and {Emery}}(1974)}]{luther-1974}
\bibinfo{author}{\bibfnamefont{A.}~\bibnamefont{{Luther}}} \bibnamefont{and}
  \bibinfo{author}{\bibfnamefont{V.~J.} \bibnamefont{{Emery}}},
  \bibinfo{journal}{Phys. Rev. Lett.} \textbf{\bibinfo{volume}{33}},
  \bibinfo{pages}{589} (\bibinfo{year}{1974}).

\bibitem[{\citenamefont{{De Grandi} and {Polkovnikov}}(2009)}]{DeGrandi2009}
\bibinfo{author}{\bibfnamefont{C.}~\bibnamefont{{De Grandi}}} \bibnamefont{and}
  \bibinfo{author}{\bibfnamefont{A.}~\bibnamefont{{Polkovnikov}}},
  \bibinfo{journal}{from ``Quantum Quenching, Annealing and Computation", Eds.
  A. Das, A. Chandra and B. K. Chakrabarti, Lect. Notes in Phys., vol. 802
  (Springer, Heidelberg 2010); arXiv:0910.2236}  (\bibinfo{year}{2009}).

\end{thebibliography}

\pagebreak

\begin{widetext}

\appendix

\section{Supplementary material for: Quantized pumping and phase diagram topology of interacting bosons}

\section{I. proof of quantized pumping by mapping to fermions}

In the main text, starting from the Sine-Gordon field theory for
the MI-HI\ transition,
\begin{eqnarray}
H_{0} &=&\frac{u}{2\pi }\int dx\Big[K\left( \partial _{x}\theta \right) ^{2}+%
\frac{1}{K}\left( \partial _{x}\phi \right) ^{2}  \notag \\
&&-g\cos (2\phi )-\lambda \sin (2\phi )\Big]  \label{H0_s}
\end{eqnarray}%
where the Luttinger parameter satisfies $1/2<K$ $<2$, we argued
that upon encircling the point $\left( g=0,\lambda =0\right) $
adiabatically, exactly one boson is transported across the system.
The argument was based on semi-classical considerations. In this
Supplementary Material, we derive this result by considering the
special \textquotedblleft Luther-Emery\textquotedblright\ point
$K=1$ \cite{luther-1974}.

At this point the hamiltonian can be rewritten in terms of non--
interacting fermions, which carry the same charge as the original
bosons. The right/left moving fermions $\psi _{\pm }$ are defined
through the bosonization formula
\begin{equation}
\psi _{\pm }=\frac{1}{\sqrt{2\pi a}}e^{i\left( \theta \pm \phi \right) }%
\text{,}
\end{equation}%
where $a$ is the microscopic cutoff of the theory. The fermionic
hamiltonian is
\begin{equation}
H_{f}=\int \frac{dk}{2\pi }\psi \left( k\right) ^{\dagger }\left[ \vec{d}%
\left( k\right) \cdot \vec{\sigma}\right] \psi \left( k\right) {^{%
\vphantom{\dagger}}}\text{.}  \label{Hf}
\end{equation}%
Here we have defined $\psi \left( k\right) ^{\dagger }\equiv
\left[ \psi
_{+}^{\dagger }\left( k\right) ,\psi _{-}^{\dagger }\left( k\right) \right] $%
, $\vec{d}\left( k\right) \equiv \left( \lambda ,g,uk\right) $, and $\vec{%
\sigma}$ are Pauli matrices. This hamiltonian is identical to the
low energy limit of the model given by Thouless in Ref.
\cite{Thouless1983}, describing spinless fermions on a
one-dimensional lattice at half filling, with a hopping
dimerization proportional to $g$ and a staggered potential
proportional to $\lambda$. The pumped charge is obtained by
computing the Chern number, which can be expressed as an integral
over the Brillouin zone
\begin{equation}
C_{\ell }=\int_{\ell }{\frac{d{\gamma }dk}{4\pi }}\hat{d}\cdot \left( \frac{%
\partial \hat{d}}{\partial \gamma }\times \frac{\partial \hat{d}}{\partial k}%
\right)   \label{Cl}
\end{equation}%
where $\hat{d}\equiv \vec{d}/\left\vert \vec{d}\right\vert $ and
${\gamma \in }\left[ -\pi,\pi \right] $ parametrizes the loop
$\ell =A,B$ (see Fig. 1a in the main text). The integral
(\ref{Cl}) measures the number of times the mapping $\hat{d}\left(
\gamma ,k\right) $ covers the unit sphere, and hence it is
quantized. Note that it is, in general, impossible to compute the
Chern number from the continuum hamiltonian (\ref{Hf}), since the
latter is accurate only in a small range of momenta up to a cutoff
$\Lambda \sim \frac{1}{a}$, while the Chern number requires
knowledge of the hamiltonian in the entire Brillouin zone.

In order to extract the Chern number from our low energy model, we
subtract the Chern number of loop B which does not enclose the
critical point from that of the nearby loop A which does enclose
it. This procedure removes the sensitivity to momenta near
$\Lambda $. In Fig. \ref{fig:d_k} we show schematically
$\hat{d}\left( \gamma ,k\right) $ for the two loops A and B. In
order to obtain $C_A-C_B$, we replace $k \rightarrow -k$ for loop
$B$ (which changes the sign of the integrand in Eq. \ref{Cl}), and
perform the integral over the combined $k$ range for both $C_A$
and $C_B$. Note that the vector $\hat{d}$ satisfied periodic
boundary conditions at the ends of this region, and therefore
$C_A-C_B$ is guaranteed to be an integer. From the figure, and
from the geometric interpretation of $C_{A,B}$ as the area covered
by $\hat{d}\left( \gamma ,k\right) $ on the unit sphere, one can
see that $C_{A}-C_{B}=1$. Therefore, loop A corresponds to pumping
of exactly one more charge than loop B. The fact that loop B is
contractible to a point without ever closing a gap implies that
$C_{B}=0$ and hence $C_{A}=1$.

The Chern number can be generalized to interacting systems by
writing it in
terms of the dependence of the many-body ground state on a phase twist $%
\varphi$ in the periodic boundary conditions\cite{NiuThouless}.
From this construction, it is clear that the value of the
quantized charge along the loop $A$ cannot change even when we
move away from the free Fermion
``Luther-Emery'' point $K=1$. We conclude that any path 
which can be deformed adiabatically to a cycle which encircles the
MI--HI critical point in the $(g,\lambda)$ space entails pumping
of a single boson across the system.

\begin{figure*}[t]
\begin{center}
\includegraphics[width=0.65\textwidth]{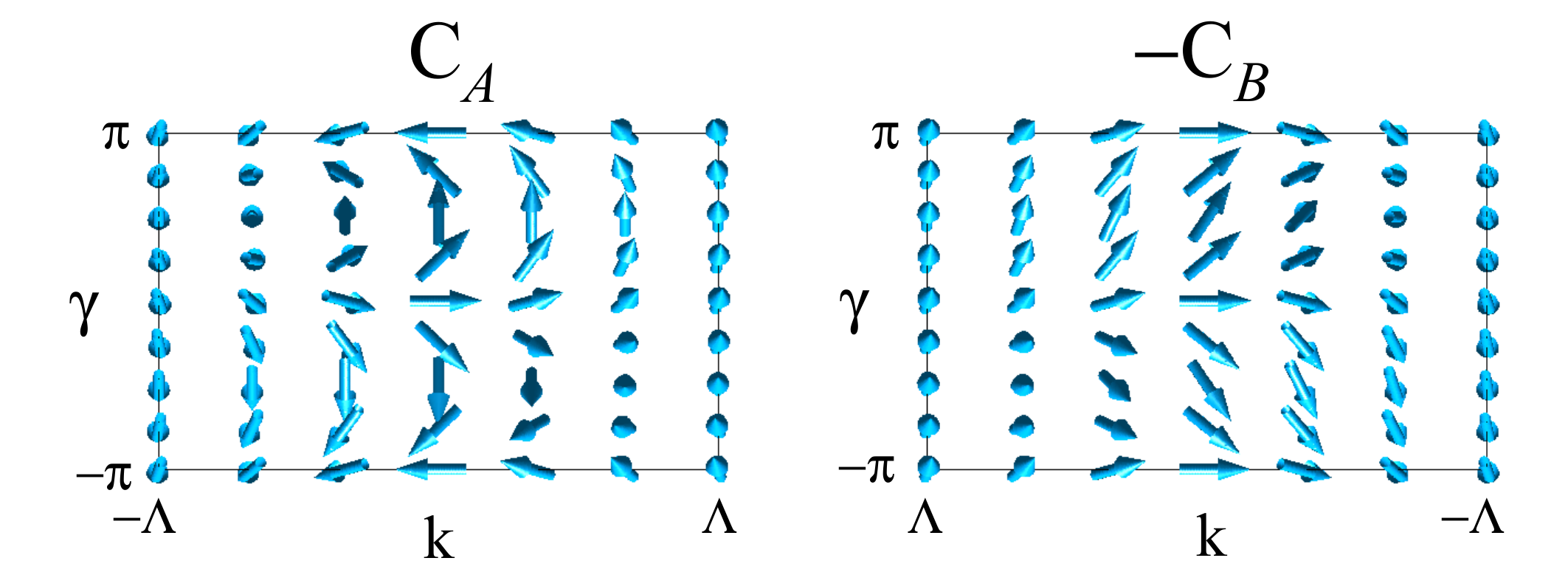}
\end{center}
\par
\vspace{-.3in} \caption{Direction of the unit vector $\hat{d}$
(see Eq. \ref{Hf}) as a function of momentum $k$ and the loop
parameter $\gamma$. Left: loop A (shown in Fig. 1a in the main
text), which encloses the critical point; right: loop B, which
does not enclose the critical point.} \label{fig:d_k}
\end{figure*}

\begin{figure}[b]
\begin{center}
\includegraphics[width=0.4\textwidth]{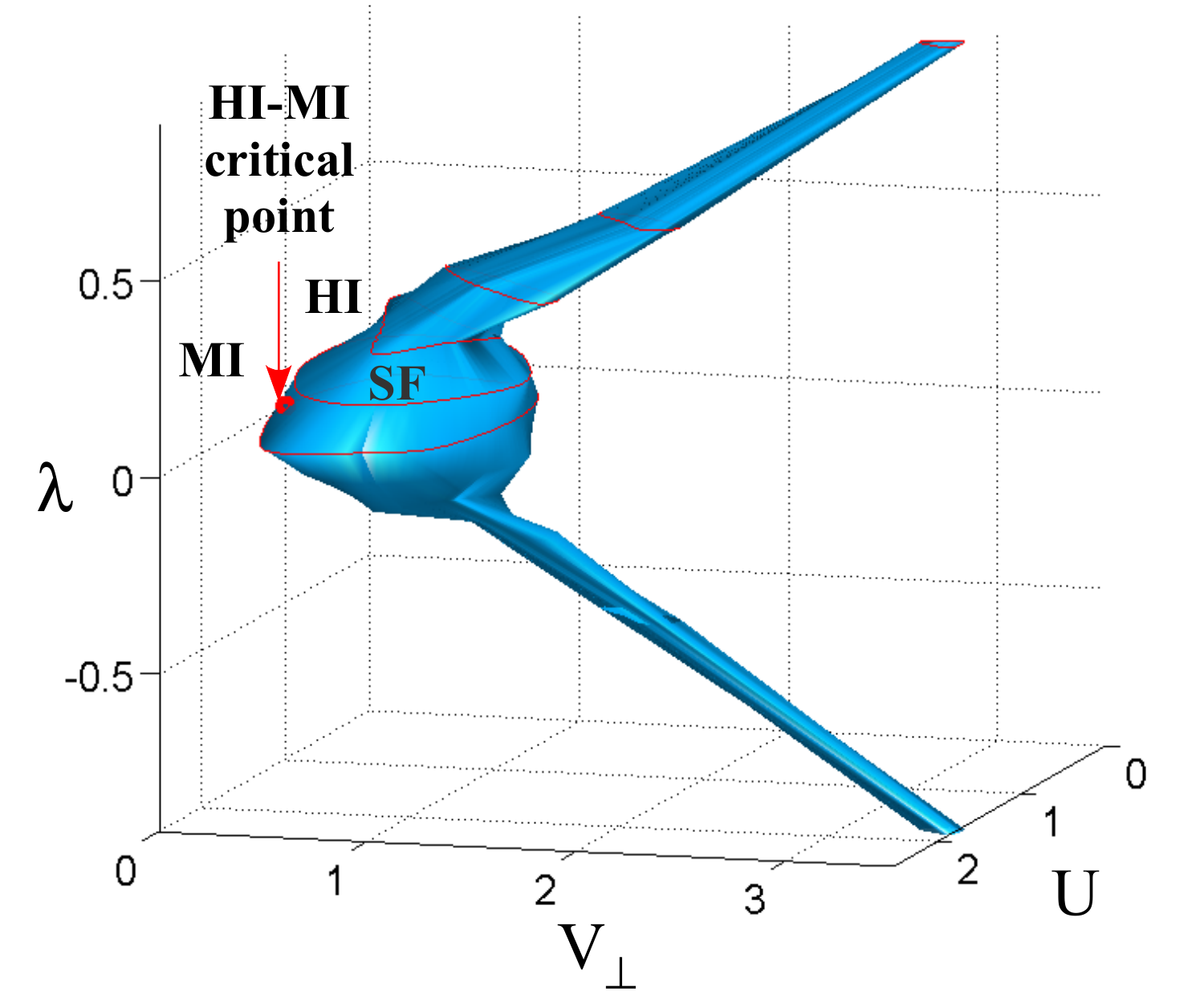}
\end{center}
\par
\vspace{-.3in} \caption{Phase diagram of the spin-1 two--leg
ladder defined in Eq. (6) (main text), as a function of $U$,
$V_\perp$, and $\protect\lambda$ (same as Fig. 3 in the main text,
but extended up to $\lambda=1$).} \label{fig:dmrg2}
\end{figure}

\section{II. Adiabaticity condition for pumping}

The adiabaticity condition required for quantization of the
pumping can be obtained from adiabatic perturbation theory (see
Ref. \cite{DeGrandi2009}). Given a time dependent Hamiltonian, we
expand the evolving wave-function in the instantaneous eigenstates
of $H(t)$ \be \ket{\psi(t)}=\sum_n a_n(t)\ket{\phi_n(t)}, \ee
where $H(t)\ket{\phi_n(t)}=E_n(t)\ket{\phi_n(t)}$. It is then
useful to define $\a_n(t)$ through the gauge transformation
$a_n(t)=\a_n(t)\exp\left(-i\Theta_n(t)\right)$, where
$\Theta_n(t)\equiv\int_0^tE_n(t')dt'$.

Let us now assume that initially only the ground state is occupied
such that $\a_n(0)=\d_{n0}$. The rate of change of the the
excitation amplitudes is then \be
\dot{\a}_n=-\bra{n}\partial_t\ket{0}e^{i\Theta_n(t)}=-{\bra{n}\partial_t
H\ket{0}\over E_n(t)-E_0(t)}e^{i\Theta_n(t)} \label{eq:an} \ee

We are interested in excitations above the gap, so that
$E_n(t)-E_0(t)\approx\D$. In addition, $\partial_t H=g\dot\chi\int
dx\sin\left(2\phi+\chi(t)\right)$. We shall denote by $C_{0n}$ the
time independent matrix element $\bra{n}\int
dx\sin\left(2\phi+\chi(t)\right)\ket{0}$. Plugging this into Eq.
(\ref{eq:an}) and integrating over time we have: \be
\a_n(t)=-{C_{n0}g\over\D}\int_{t_0}^t dt'\dot{\chi}e^{i\D t'} \ee
To compute this amplitude we should specify how the time
dependence is turned on and off. Since we are interested in the
adiabatic limit it is natural to do this in a smooth way. We take
$\chi(t)=\pi\tanh(t/\tau)$.

\section{III. Phase diagram at higher values of $\lambda$}

In order to verify that the gapless superfluid region in the phase
diagram (Fig. 2 in the main text) does not terminate, we have
extended the calculation to larger $\lambda$ values. Fig.
\ref{fig:dmrg2} shows the phase diagram of the effective spin
model (Eq. 6 in the main text) with $V_\perp=2t_\perp$, extended
up to $\lambda=\pm 1$. Consistently with the charge pumping
argument, we found that the superfluid ``horns'' extend in the
$\pm \lambda$ direction and persist to the largest $\lambda$ we
have examined. Further calculations show that the superfluid
region extends to at least $\lambda=\pm 2$ (not shown).
Ultimately, the superfluid region can either extend to $\lambda
\rightarrow \pm \infty$, or collapse onto itself to form a
donut-like shape.

\end{widetext}

\end{document}